\documentclass[fleqn,twoside]{article}
\usepackage[headings]{espcrc2}

\readRCS
$Id: espcrc2.tex,v 1.2 2004/02/24 11:22:11 spepping Exp $
\usepackage{graphicx}
\usepackage[figuresright]{rotating}

\newcommand{\AmS}{{\protect\the\textfont2
  A\kern-.1667em\lower.5ex\hbox{M}\kern-.125emS}}

\title{Recent Developments in Parallelization of the Multidimensional 
       Integration Package DICE}

\author{F. Yuasa\address[KEK]{High Energy Accelerator Research Organization, KEK,
        1-1 OHO Tsukuba, Ibaraki  305-0801, Japan}
        \thanks{{\it E-mail address} : fukuko.yuasa@kek.jp},
        K. Tobimatsu\address{Kogakuin Univ., 1-24-2 Nishi-Shinjuku,
        Shinjuku-ku, Tokyo 163-8677, Japan }
        and
        S. Kawabata\addressmark[KEK]}
       
\runtitle{Recent Developments in Parallelization of the Multidimensional 
       Integration Package DICE}
\runauthor{F.Yuasa}

\begin{document}

\begin{abstract}
DICE is a general purpose multidimensional numerical integration package. 
There can be two ways in the parallelization of DICE, 
``distributing random numbers into workers'' 
and ``distributing hypercubes into workers''. 
Furthermore, there can be the combination of both ways.
So far, we had developed the parallelization code using the former
way and reported it in ACAT2002 in Moscow.
Here, we will present the recent developments of parallelized DICE 
in the latter way as the 2nd stage of our parallelization activities.
\vspace{1pc}
\end{abstract}

\maketitle

\section{Introduction}
Recently it is not rare to calculate the cross sections of the physics 
process with over 6 final state particles in the tree level. 
In such calculations, there may appear singularities close to diagonal 
integral 
region and sometimes it is very difficult to find a good set of variable 
transformations to get rid of the singularities. For the one-loop and beyond 
the one-loop physics processes, when we try to carry out the loop calculation
only in the numerical approach, we need several multidimensional integration 
packages or another integration method to compare the numerical results to check them.
For such a request, DICE has been developed by K.Tobimatsu and S.Kawabata.
It is a general purpose multidimensional numerical integration package.

\subsection{The non-parallelized version of DICE}
The first version of DICE\cite{DICE1} appeared in 1992 and is a scalar 
program code. In DICE, the integral region is divided into 
$2^{N_{dim}}$ hypercubes repeatedly according to the division condition. 
To evaluate the integral and its variance in each hypercube, DICE tries
two kinds of sampling method, a regular sampling and a random sampling
as:

\begin{enumerate}
\item{Apply regular sampling and evaluate the contribution. \\
And then check the division condition is satisfied or not. }
\item{Apply 1st random sampling and evaluate the contribution. \\
And then check the division condition is satisfied or not. }
\item{Apply 2nd random sampling and evaluate the contribution.}
\end{enumerate}

For an integrand with singularities the number of above repetitions 
becomes huge so rapidly and the calculation time becomes a long time.
To reduce the calculation time, the vectorized version of 
DICE (DICE 1.3Vh\cite{DICE2}) has been developed  in 1998 for vector 
machines. In the vector program code, the concept of workers and 
the queuing mechanism are introduced. 
This vectorized DICE has succeeded the reduction of the calculation 
time for the integration even when the integrand  has strong 
singularities. 

Today, however, the vector processor architecture machines have dropped off
and instead the parallel processor architecture machines become common in 
the field of High Energy Physics. Moreover, the cost effective  
PC clusters running Linux with distributed memory or shared memory are widely 
spread. Thanks to this rapid rise of PC clusters with the Parallel library 
such as MPI\cite{MPI} or with OpenMP\cite{OPENMP}, the parallel programming 
is very familiar to us.

\section{Parallelization}

\subsection{Profile of DICE}
To get a good efficiency in the parallelization, it is important to know
which routines are time-consuming. UNIX command {\tt{gprof}} is a useful 
tool to know it. 
In Table~\ref{table:1}, an example output of {\tt{gprof}} command for 
the calculation of the integration by the non-parallelized DICE. 
This calculation is done on the Alpha 21264 processor (700 MHz clock speed)
machine running Linux and the compiler used is  Compaq Fortran.
In Table ~\ref{table:1}, the most time-consuming routine is {\tt{elwks}} 
and is called in {\tt{func}}. In {\tt{elwk}} and {\tt{func}} the 
integrand function is given.
The subroutine {\tt{func}} is called repeatedly in {\tt{regular}},
{\tt{random1}} and {\tt{random2}} to evaluate the integrand. 
Here, {\tt{vbrndm}} is a routine to generate
random numbers and is called in both {\tt{random1}} and {\tt{random2}}.

In summary, it is expected that distributing the calculations 
in {\tt{random1}} and {\tt{random2}} into workers (processors) may be 
efficient to reduce the calculation time. 

\begin{table*}[htb]
\caption{{\tt{gprof}} output : Flat profile of non-parallelized DICE}
\label{table:1}
\newcommand{\m}{\hphantom{$-$}}
\newcommand{\cc}[1]{\multicolumn{1}{c}{#1}}
\renewcommand{\tabcolsep}{1.3pc} 
\renewcommand{\arraystretch}{1.2} 
\begin{tabular}{lllllll}
\hline
Time & Cumulative & Self & Calls & Self & Total & Name of \\
\% & time [sec] & time [sec] &  & [$\mu$/call] & & routines \\ \hline
82.95 & 7.60 & 7.60 & 26214 & 0.29 & 0.29 & \tt{elwks\_} \\ 
12.41 & 8.73 & 1.14 & 26214 & 0.04 & 0.33 & \tt{func\_} \\
2.52  & 8.96 & 0.23 & 3072  & 0.08 & 0.08 & \tt{vbrndm\_} \\
0.93  & 9.05 & 0.08 & 1536  & 0.06 & 2.80 & \tt{randm2\_} \\
0.92  & 9.13 & 0.08 & 1536  & 0.05 & 2.79 & \tt{randm1\_} \\
0.13  & 9.14 & 0.01 & 1638  & 0.01 & 0.34 & \tt{regular\_} \\ \hline
\end{tabular}\\[2pt]
This calculation of the integration is done on the Alpha 21264/700MHz machine by the 
Compaq Fortran for Linux.
Total CPU time required was 9.16 sec in total. This integration was done with expected error = 10\%.
\end{table*}

\subsection{Algorithm}
For the integrand with strong singularities, the region is divided into 
a large number of hypercubes and the total number of random numbers are 
required to get the integral results with the requested errors. 
Therefore, there can be two ways in the parallelization, the way of 
distributing random numbers and the way of distributing hypercubes to 
workers. 

\begin{figure}[htb]
\begin{center}
\includegraphics[width=5cm]{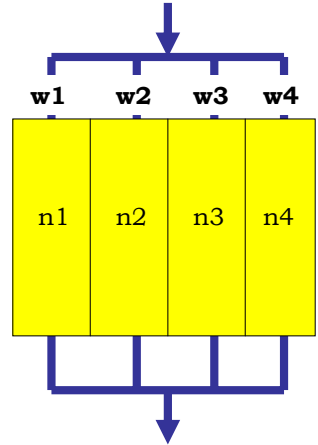}
\caption{Schematic view of the algorithm of parallelized DICE with 
distributing the random numbers into workers, w1, w2, w3 and w4, for example. 
Each worker is responsible for the part of random numbers in the routines, 
{\tt{random1}} and {\tt{random2}}. The total number of random numbers is 
the sum of random numbers treated in each worker as 
$N_{total} = n1 + n2 + n3 + n4$.}
\label{fig:fig1}
\end{center}
\end{figure}

As the 1st step we have started the parallelization of DICE with the former way,
distributing random numbers. The schematic view of the algorithm 
with the former way is shown in Fig.~\ref{fig:fig1}.
There, it is shown how random numbers are distributed 
into workers in {\tt{random1}} and {\tt{random2}}.
The merit of this approach is not only that the algorithm is very simple 
as shown in  Fig.~\ref{fig:fig1} but also that the overhead due to data transfer or load 
unbalancing among workers.
The efficiency of this parallelization have showed very good performance. 
The result of the efficiency measurement by this parallelization way was
presented at ACAT2002 at MOSCOW~\cite{ACAT2002}. 

As the 2nd step, here in this paper, we present the parallelization with 
the latter way, distributing hypercubes into workers. As the 3rd step, the final step, we have a plan of the combination of both ways.

\section{Implementation}

\begin{figure}[htb]
\includegraphics[width=7cm]{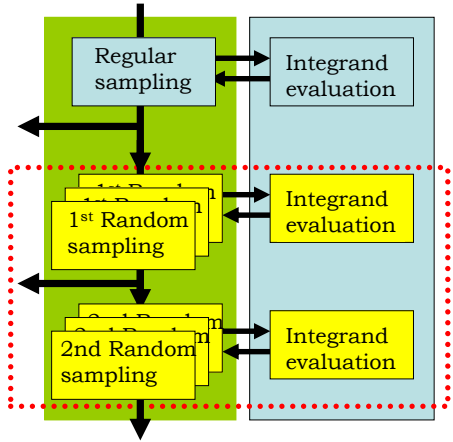}
\caption{Schematic view of the algorithm of Parallelized DICE. Each worker
is responsible for the part of hypercubes.}
\label{fig:fig2}
\end{figure}

In this parallelization,  hypercubes are distributed into workers. After 
the evaluation, the results are gathered to the root process (for example, 
worker 1). And then the root process scattered the results 
to all workers. In Fig.~\ref{fig:fig2}, a schematic view how calculations 
are distributed into workers is shown.

In our implementation we use Fortran compiler since DICE is written 
in {\tt{Fortran}} and we chose {\tt{MPI}}\cite{MPI}
as the parallel library.

\section{Efficiency Measurement}

\subsection{Measurement Environment}

In Table~\ref{table:2}, the measurement environment is shown.
There, MPI bandwidth was measured  and was 56.79 MB/s. 
It is measured by a simple ping-pong program using MPI send-receive functions. 
The size of the transferred data is 1 MB and the figure is an average 
by 10 times measurement. In all measurements we use a cheap Gigabit Ethernet 
switch and 
it can be said that the switch showed a reasonable performance.

\begin{table}[h]
\caption{Measurement Environment : PC cluster}
\label{table:2}
\begin{tabular}{|l|l|}
\hline
CPU & Xeon dual 3.06GHz \\ \hline
\# of systems & 8 systems\\ \hline
Memory & 2 [GB]\\ \hline
Switch & 10/100 /1000 switch \\ \hline
Compiler & /usr/local/mpich-intel81/ \\ 
 & /bin/mpif77 \\ \hline
MPI Bandwidth & 56.79 [MB/s] in average \\ \hline
\end{tabular}\\[2pt]
\end{table}

\subsection{Example Physics process: $e^+e^- \rightarrow \mu^+ \mu^- \gamma$}

We choose the radiative muon pair production as an example physics process 
to measure the efficiency of the parallelization. This physics process 
is a realistic example and it must be a good example since there are several 
singularities in the calculation of the cross section. 
That is, there are the mass singularities in the initial electron and 
positron, those for the final muons, the s-channel singularities caused by 
nearly on-shell s-channel photon by a hard photon emission from the initial 
electron or positron, and the singularity by the infra-red divergence of a 
real photon which is regularized by introducing a cutoff for the photon 
energy $k_{c}$.

In Table~\ref{table:3}, we summarized the parameters in the measurements. 
There we used the naive kinematics which means the kinematics without
finding a good set of variables. That means  there still exist several 
singularities in the integrand. The details of the naive kinematics 
are shown in Ref\cite{DICE2}.
As a matter of course, the studies of the kinematics for this example process 
have been well done and we should add that there exists one program code of 
the kinematics with a well selected set of variables to avoid the strong 
singularities\cite{SHIMIZU}.  

\begin{table}[t]
\caption{Summary of the parameters in the measurements}
\label{table:3}
\begin{tabular}{|l|l|}
\hline
Physics process & $e^+e^- \rightarrow \mu^+ \mu^- \gamma$ \\ \hline
$E_{CM}$ &  70 [GeV] \\ \hline
$k_{c}$  & 100 [MeV] \\ \hline
kinematics& naive kinematics\\ \hline
\# of dimensions& 4 \\ \hline
\# of random numbers & \\
in each hypercube & 100 \\ \hline
Max. \# of workers & 8 \\ \hline
\end{tabular}\\[2pt]
\end{table}

\subsection{The Wall-Clock Time Measurement }
Roughly speaking, in the parallel calculation the CPU time in 
each worker must decrease $1/2$, $1/4$, and $1/8$ when the 
number of workers increases as 2, 4, and 8.
This is the basic check whether the parallel code runs well.
Actually the reduction rate of the wall-clock time is a more important
measure than the reduction rate of CPU time to see how efficient
the parallel code is. 

In Table~\ref{table:4} and ~\ref{table:5}, the measured wall-clock time 
are shown for the calculations of the cross section with 
requested errors which are 1\% and 2\% respectively. 
In both Tables, the wall-clock time 
becomes shorter and  both reduction rates are in the same manner.
However, the reduction rate is not good enough when the number of workers
increases. 

\begin{table*}[htb]
\caption{Efficiency of the Parallelized DICE : The calculation of the 
cross section with the  error = 1.97\%. $\sigma = (2.5824 \pm 0.0508) \times 10^{-2}$ [nb
] }
\label{table:4}
\newcommand{\m}{\hphantom{$-$}}
\newcommand{\cc}[1]{\multicolumn{1}{c}{#1}}
\renewcommand{\tabcolsep}{1.3pc} 
\renewcommand{\arraystretch}{1.2} 
\begin{tabular}{@{}lllll}
\hline
\# of Processors & CPU time & wall-clock time & Reduction rate: & Reduction rate:\\
(Workers) & [sec] & [sec] & CPU time & wall-clock time \\ \hline
1 & 8882.03 & 8894 & 1.00 & 1.00 \\ 
2 & 5179.62 & 6178 & 0.58 & 0.69 \\
4 & 3308.92 & 5011 & 0.37 & 0.56 \\
8 & 2394.34 & 4784 & 0.27 & 0.54 \\ \hline
\end{tabular}\\[2pt]
Xeon 3.06 GHz, non-parallelized DICE, required CPU time is 8704.84 sec.
\end{table*}

\begin{table*}[htb]
\caption{Efficiency of the Parallelized DICE : The calculation of the 
cross section with error = 0.91\%. $\sigma = (2.8307 \pm 0.0259) \times 10^{-2}$ [nb
] }
\label{table:5}
\newcommand{\m}{\hphantom{$-$}}
\newcommand{\cc}[1]{\multicolumn{1}{c}{#1}}
\renewcommand{\tabcolsep}{1.3pc} 
\renewcommand{\arraystretch}{1.2} 
\begin{tabular}{@{}lllll}
\hline
\# of Processors & CPU time & wall-clock time & Reduction rate: & Reduction rate:\\
(Workers) & [sec] & [sec] & CPU time & wall-clock time \\ \hline
1 & 186682.83 & 197444 & 1.00 & 1.00 \\
2 & 109401.92 & 134377 & 0.59 & 0.68 \\
4 &  69676.01 & 108234 & 0.37 & 0.55 \\
8 &  51056.55 & 103126 & 0.27 & 0.52 \\ \hline
\end{tabular}\\[2pt]
Xeon 3.06 GHz, non-parallelized DICE, required CPU time is 183884.34 sec.
\end{table*}

\section{Summary and Outlook}

We presented recent developments on the parallelization of DICE.
The ongoing work is in the 2nd stage of our activities for the parallelization.
Efficiency of the current parallel code has been evaluated for the example
physics process $e^+e^- \rightarrow \mu^+ \mu^- \gamma$ with naive kinematics.
For this process, the wall-clock time was actually reduced with the current 
parallel code but the reduction rate is not satisfactory when the number of 
workers increases. So there is still some work remained to optimize the 
current parallel code further.

Our current code is based on the vectorized DICE,
DICE 1.3vh, and in it the load balancing mechanism between workers is not 
included. We believe that further more reduction of the wall-clock time will 
be possible with applying the load balancing mechanism to our current code.

The main goal of all efforts is 
the parallelization using the combination of both distribution ways.
After including the load balancing mechanism we will be able to enter the 3rd 
stage of the parallelization. 

\par
~\\
\noindent
{\large{\bf{Acknowledgments}}}\\
We wish to thank the members of MINAMI-TATEYA collaboration for discussions.
We also wish to thank Prof. Y.Shimizu for encouragements.
This work was supported in part by the Grants-in-Aid (No. 17340085)
 of Monbukagaku-sho, Japan.

\end{document}